\documentclass[aps,pre,preprint,floatfix]{revtex4-1}
\usepackage[utf8]{inputenc}
\usepackage{amsmath}
\usepackage{amsfonts}
\usepackage{amssymb}
\usepackage{graphicx}
\usepackage{xcolor}
\usepackage{multirow}
\usepackage[alsoload=synchem]{siunitx}

\begin{document}
\title{Phase Controllable Josephson Junctions for Cryogenic Memory}

\author{Alexander E Madden, Joshua C Willard, Reza Loloee, and Norman O Birge}
\email{birge@pa.msu.edu}
\affiliation{Department of Physics \& Astronomy, Michigan State University, East Lansing, Michigan 48824-2320, USA}

\begin{abstract}
Josephson junctions containing ferromagnetic layers have generated interest for application in cryogenic memory. In a junction containing both a magnetically hard fixed layer and soft free layer with carefully chosen thicknesses, the ground-state phase difference of the junction can be controllably switched between 0 and $\pi$ by changing the relative orientation of the two ferromagnetic layers from antiparallel to parallel. This phase switching has been observed in junctions using Ni fixed layers and NiFe free layers. We present phase-sensitive measurements of such junctions in low-inductance symmetric SQUID loops which simplify analysis relative to our previous work. We confirm controllable $0-\pi$ switching in junctions with \SI{2.0}{\nano\meter} Ni fixed layers and \SI{1.25}{\nano\meter} NiFe free layers across multiple devices and using two SQUID designs, expanding the phase diagram of known thicknesses that permit phase control.
\end{abstract}

\maketitle

\section{Introduction}
Experimental studies of ferromagnetic Josephson junctions have shown incredible promise over the past two decades since the first demonstrations of so-called $\pi$-junctions \cite{Ryazanov2001,Kontos2002}. In the conventional model of a Josephson junction, the current through the junction depends on the phase difference $\phi$ of the order parameter across the junction as $I_s=I_c \sin\phi$, where $I_c$ is the junction's critical current above which the junction develops a voltage. Such junctions have minimal energy when $\phi=0$. Richer behavior is observed in S/F/S junctions where the barrier layer is replaced with a ferromagnet \cite{Golubov2004,Buzdin_SFReview2005}. A Cooper pair in an s-wave superconductor consists of two paired electrons with equal and opposite momenta and opposite spins. When two such paired electrons enter the ferromagnetic layer, they enter different spin bands and acquire a net center-of-mass momentum $\pm \hbar Q=\hbar(k_F^\uparrow - k_F^\downarrow)$ with the Fermi momenta in the spin up and down bands given by $k_F^\uparrow$ and  $k_F^\downarrow$ respectively \cite{DemlerArnoldBeasly}. This momentum manifests as an oscillation in the pair correlation function as it decays into the magnetic layer. For certain thicknesses of ferromagnet, this oscillation can lead to minimization of the junction's energy at $\phi=\pi$ instead of the usual zero \cite{Buzdin1982,Buzdin1991}.

Junctions containing multiple ferromagnetic layers exhibit even richer behavior.  It was predicted early on that a single junction containing two ferromagnetic layers could be switched between a 0-state and a $\pi$-state by changing the relative orientation of the two magnetizations \cite{Bergeret2001PRL,Krivoruchko2001,Golubov2002,Barash2002}. Among those works, the one by Golubov, Kupriyanov, and Fominov \cite{Golubov2002} is of particular relevance to our experiments, since it addresses the case of an S/F1/F2/S junction with thick superconducting electrodes and ferromagnetic materials with exchange energy larger than the superconducting gap.  Further theoretical work was presented on such systems in both the clean and dirty limits \cite{Blanter2004,Pajovic2006,Crouzy2007}.  The physical explanation of the effect is straightforward: when the magnetizations are parallel the two ferromagnetic layers function as one effective thicker magnetic layer, and the pair correlation function accumulates a phase $\phi_P = d_{F1}Q_1+d_{F2}Q_2$ as it traverses both layers.  Alternatively, if the layers are antiparallel then the phase accumulation through the magnetic layers is $\phi_{AP} = d_{F1}Q_1-d_{F2}Q_2$.  One can choose the thicknesses so that these two situations produce different phase states of the junction:  for example, if the first magnetic layer has a thickness $d_{F1}$ close to its $0-\pi$ transition thickness and the second has a thickness $d_{F2}$ less than its transition thickness, then the parallel case produces a junction in the $\pi$-state, while the antiparallel case produces the 0-state \cite{Golubov2002, Blanter2004,Pajovic2006,Crouzy2007}.

The first experimental work to address Josephson junctions with two ferromagnetic layers was by Bell \textit{et al.} \cite{Bell2004}.  Those authors employed a Co/Cu/Permalloy ``pseudo-spin-valve'' structure (Permalloy = Ni$_{80}$Fe$_{20}$) inside their junctions, which enabled them to switch the magnetization direction of the magnetically soft Permalloy layer without switching the magnetically harder Co layer.  Although the experiments were sensitive only to the magnitude of the critical current, those authors did speculate that it should also be possible to control the junction phase state. Furthermore, Bell \textit{et al.} proposed that such controllable Josephson junctions could be used as memory elements in a cryogenic memory.

Research on cryogenic memory has been underway for several decades \cite{Hebard1978,Uehara1981,Faris1980,Miyahara1984,Miyahara1985}. The standard scheme of storing flux quanta in superconducting loops does not scale well to small sizes \cite{Nagasawa1995,RYAZANOV201235}. Hence, many groups have searched for alternative memory technologies including various forms of magnetic memory  \cite{Beasley1997,Johnson1999,VanDuzer2005,Mannhart2006}. This research has surged in recent years   \cite{RYAZANOV201235,Mukhanov2012,Koelle2013,Buhrman2014,Soloviev2014,Golod2015,Bakurskiy2016,Gibson2017,Bakurskiy2018}, partly to address the need for energy-efficient large-scale computing \cite{Holmes2013,Soloviev2017_review}. Several groups have explored using pseudo-spin-valve junctions for this purpose \cite{Baek2014,Qader2014,Gingrich2016,Niedzielski2018}.
Our work is largely motivated by the Josephson Magnetic Random Access Memory (JMRAM) architecture recently demonstrated \cite{Dayton2018} by Northrop Grumman Corporation. In this design, the bit of the memory device is represented by the phase state of an S/F1/F2/S junction in a SQUID loop with two standard S/I/S junctions. The magnetic junction is designed to have a much higher critical current than the S/I/S junctions, allowing it to stay in the superconducting state during the read operation. The read speed is then governed by the much larger $I_cR_n$ of the SIS junction, allowing faster readout and a stronger signal than a voltage measurement on the ferromagnetic junction.

Our previous study of Ni/NiFe spin-valve junctions included the first phase-sensitive demonstration of $0-\pi$ switching in a ferromagnetic Josephson junction \cite{Gingrich2016}. In that experiment, junctions with a \SI{1.5}{\nano\meter} NiFe free layer and \SI{1.2}{\nano\meter} Ni fixed layer were used in an asymmetric SQUID with different inductances for the two arms to indentify which junction was switching when a phase change was observed. Although it is difficult to control nickel's magnetic state due to it's multidomain structure, our group and others have found it to be one of the best ferromagnetic materials to pass large supercurrents \cite{Baek2017}. SQUID critical current oscillations were measured as a function of the flux through the loop while a separate external field was applied to switch the state of the junctions. When the junctions switched, the SQUID oscillations showed a phase shift as expected, as well as a change in amplitude. Unfortunately the SQUID oscillations were fairly complicated, showing an asymmetric ``ratchet'' shape due to the large inductances of the arms and asymmetric design. This asymmetry also caused an additional shift separating the maxima of $I_c^+$ and $I_c^-$, the critical currents extracted from the positive and negative halves of the IV curves. Preliminary fits to the data from that study often gave multiple possible values for the phase shift because this asymmetric offset was comparable to the period of the oscillation. It was possible to extract a unique phase shift from careful analysis of the oscillations and comparisons between states, but having the phase shift directly observable in the raw data would simplify the analysis and interpretation considerably.

In this study, two new SQUID designs were used: one similar to the previous architecture but with symmetric arms and another designed to significantly reduce the self-inductances of the SQUID arms. Both designs were intended to present more clear and definitive evidence of $0-\pi$ switching. Moving to low-inductance symmetric SQUIDs greatly simplifies the analysis and in many cases allows the phase shift to be observed directly in the critical current oscillations.

A more recent study \cite{Niedzielski2018} on single Ni/NiFe junctions identified additional material thicknesses that should support $0-\pi$ switching but requires phase-sensitive confirmation. Based on the range of NiFe thicknesses suggested by that work, this study includes phase-sensitive measurements confirming $0-\pi$ switching for \SI{2.0}{\nano\meter} Ni and \SI{1.25}{\nano\meter} NiFe. This is an important extension of the phase diagram mapping of thicknesses supporting phase control.

As an historical note, the very first S/F/S $\pi$ junctions using very weak ferromagnets were in some sense controllable because they underwent a $0-\pi$ or $\pi-0$ transition as a function of temperature \cite{Ryazanov2001,Frolov2004}.  Several other types of controllable junctions have been proposed or demonstrated.  Long S/N/S junctions can be converted to $\pi$-junctions by injecting current into the normal part of the junction \cite{Baselmans1999,Baselmans2001,Huang2002,Zaikin1998,Yip1998}.  A Zeeman (magnetic) field applied to an S/N/S junction can produce a controllable $\pi$ junction \cite{Yip2000,Heikkila2000}, as can injection of a nonequilibrium spin population into an S/N/S junction \cite{Yamashita2006}. Phase control has been demonstrated using electrostatic gating in carbon nanotube \cite{cleuziou2006,Delagrange2018} and quantum dot junctions \cite{vanDam2006,Szombati2016}. There are theoretical proposals to produce controllable $\pi$ junctions using spin-triplet superconductors and ferromagnets \cite{Kastening2006}, using a quantum-dot Josephson junction containing a molecular spin \cite{Benjamin2007}, by electrostatically gating MoS$_\textrm{2}$ monolayers \cite{Rameshti2014}, and with pinned Abrikosov vortices \cite{Mironov2017}.  The recent intense interest in topological systems will undoubtedly turn up multiple ways to realize Josephson junctions with controllable phase states \cite{Pientka2017}.
\section{Junction Fabrication}\label{sec:Fab}

\begin{figure}[ht]
\includegraphics[width=2.4in]{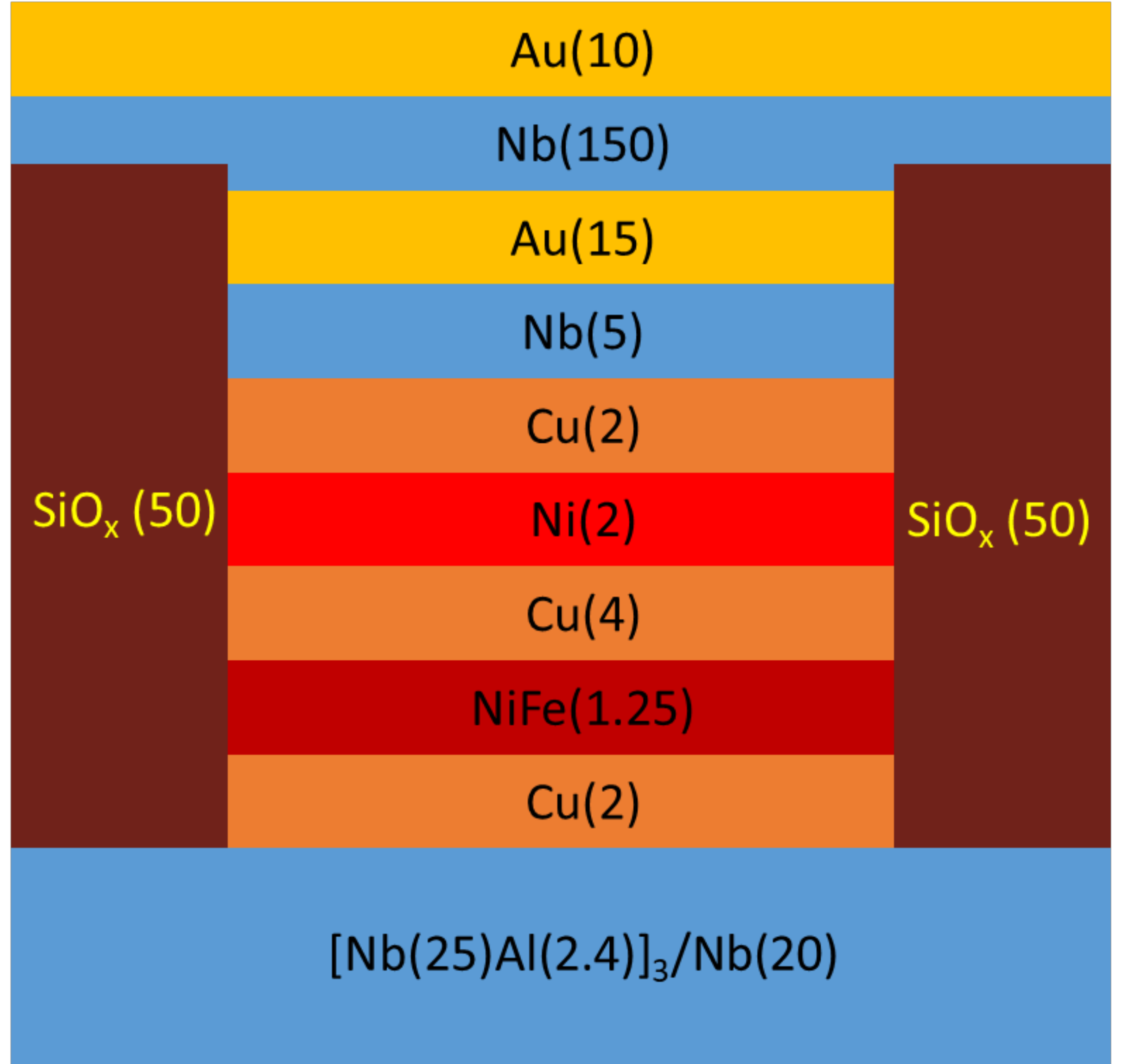}
\caption{Cartoon of materials sputtered in ferromagnetic Josephson junction. The superconducting bottom electrode consists of a niobium/aluminum base layer, which is smoother than pure niobium. A copper spacer layer is sputtered to promote sharper switching and a lower switching field in the NiFe free layer above it. Another copper spacer layer is used to decouple the two ferromagnets, and then the Ni fixed layer is sputtered. After another copper spacer, thin niobium and gold layers are deposited to protect the junction from oxidation during ion milling. An SiO$_x$ insulating layer is thermally evaporated around the pillars after ion milling before the top-lead deposition. Finally, a thick superconducting niobium top electrode is sputtered with a gold capping layer to prevent oxidation. }
\label{Stack}
\end{figure}

A cross-section of the Josephson junctions is shown in Figure \ref{Stack}. The SQUID devices were photolithographically patterned using a bilayer of LOR5B and S1813 photoresist. The bilayer gave a slight undercut in the resist after developing, which assists in liftoff by preventing metal on the sides of the deposited leads from sticking to the resist. A bottom [Nb/Al] superconducting base layer was used as previous work \cite{NiedzielskiThesis2017,Wang2012} has shown it has lower roughness than thick niobium, encouraging better growth of the ferromagnetic layers.  The  sputtered stack had the form [Nb(25)Al(2.4)]$_3$/Nb(20)/Cu(2)/NiFe(1.25)/ Cu(4)/Ni(2)/Cu(2)/Nb(5)/Au(15) with all thicknesses in nanometers and subscripts indicating repeated layers. The gold capping layer was included to prevent oxidation. Previous work by our group on normal metal buffer layers \cite{NiedzielskiThesis2017} has shown that NiFe switches more sharply and at a lower field when grown on a \SI{2}{\nano\meter} copper buffer layer, so copper was sputtered between the superconducting and ferromagnetic layers. The \SI{4}{\nano\meter} Cu spacer was used to decouple the two ferromagnetic layers so they would switch independently. Sputtering was performed at a substrate temperature between \SI{-15}{\celsius} and \SI{-30}{\celsius} in an Ar pressure of \SI{2}{\milli\torr} in a system with a base pressure of \SI{2e-8}{\torr}. Permanent magnets were placed behind the substrates during sputtering to induce magnetocrystalline anisotropy along the long axis of the junctions. The Josephson junctions were then patterned by electron beam lithography using negative resist ma-N 2401. The stack was ion milled down to the niobium base layer, leaving the full stack only under the lithographically defined junctions. After ion milling, a \SI{50}{\nano\meter} insulating SiO$_x$ layer was thermally evaporated, followed by lift-off of the e-beam resist. The chips were pressed against a copper mass coated in silver paste for heatsinking during the ion milling and SiO$_x$ deposition. The top superconducting leads were photolithographically defined and the top Nb(150)/Au(10) superconducting electrode was sputtered. 

\section{SQUID Design}
\begin{figure}[ht]
\includegraphics[width=3.4in]{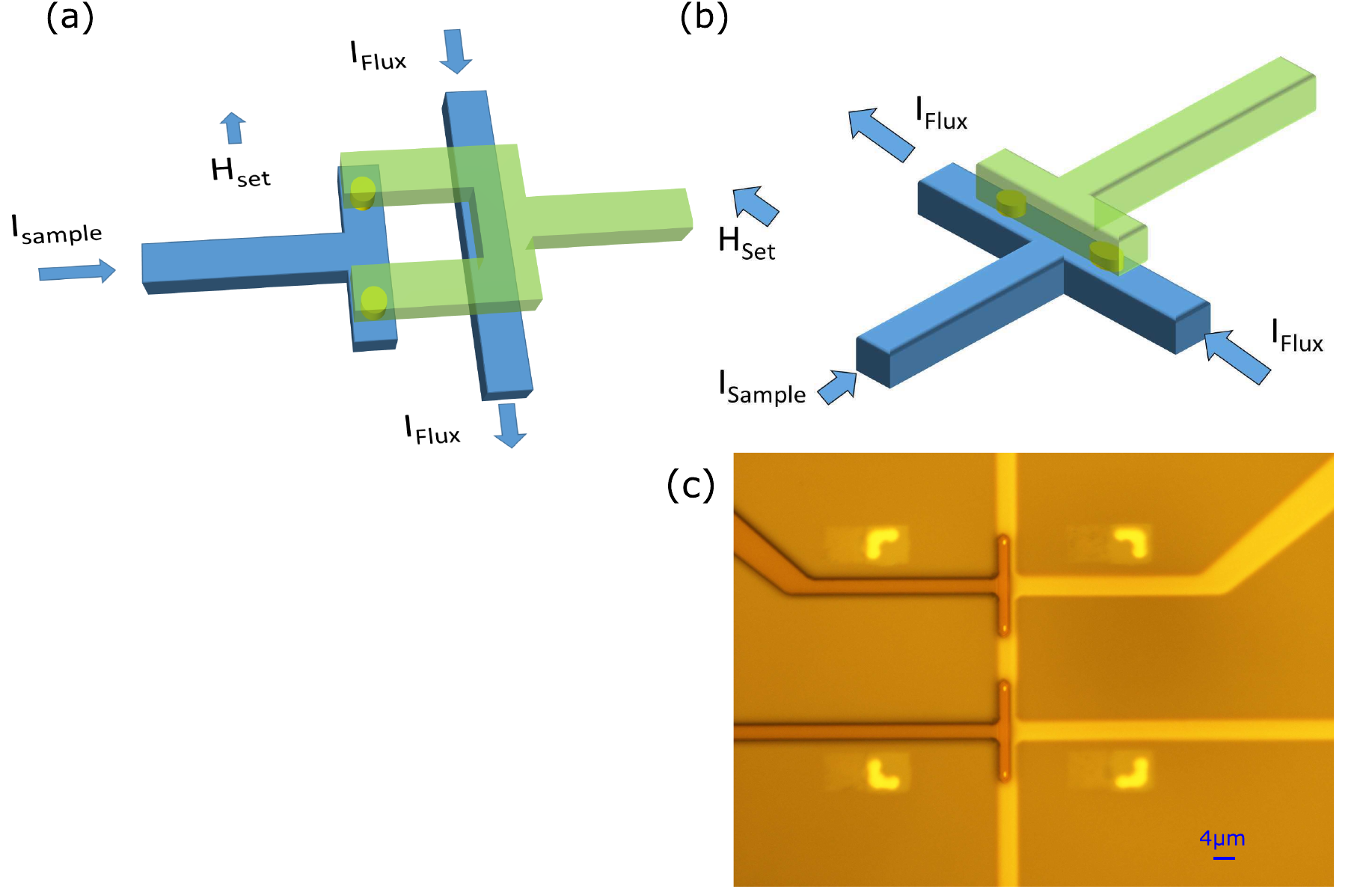}
\caption{Cartoon of different SQUID geometries used. (a) Pitchfork SQUID. Independent flux line below plane of top-lead loop. (b) Ultra-low inductance (ULI) SQUID. Flux line shares common bottom electrode with sample current. (c) Optical microscope image with approximate scale bar of two Ultra-Low Inductance SQUIDs taken after top-lead photolithography. The bottom lead including the vertical flux line and horizontal current leads appears as a lighter metallic color, and the darker color shows where the top leads will be deposited. Josephson junctions are visible as light points where the top and bottom leads overlap.}
\label{Cartoon}
\end{figure}

Two patterns of SQUIDs, shown in Figure \ref{Cartoon}, were studied in this work. The first was a symmetric ``pitchfork'' design similar to the asymmetric design used in our previous work. The sample current is injected through the bottom lead, runs through the junctions in the SQUID, and returns via the top lead. The flux threading the SQUID is controlled by an independent current line on the chip running below the SQUID. In our previous phase-sensitive study, the difference in inductances between the two arms of the SQUID caused an asymmetry in the SQUID critical current oscillations which made it more difficult to identify the phase of the oscillation. A symmetric design is used here to simplify this stage of the analysis.

The second design studied is our ``Ultra Low Inductance'' (ULI) design shown in Figure \ref{Cartoon}(b).  The sample current and flux current are injected via perpendicular wires on a common bottom electrode. An optical microscope image of this design taken before top-lead sputtering is shown in Figure \ref{Cartoon}(c). The junctions are patterned symmetrically with respect to the measurement current lead, and the top half of the SQUID lies directly above the flux line. This produces a SQUID loop with a cross-section perpendicular to the surface of the chip and a thickness governed only by the thickness of the thin insulating layer between the top and bottom leads. Because the junctions are placed symmetrically with respect to the bottom measurement current lead, the current through the sample couples no additional flux into the SQUID loop when the junction critical currents are equal. The flux current is provided using a Yokogawa current supply with a floating ground to minimize crosstalk between the measurement and flux currents. This design, as well as the pitchfork design, has the same nominal inductances for both arms. The lower inductances also reduce the relative shift between $I_c^+$ and $I_c^-$ that causes the distinctive ``ratchet'' shape in the SQUID oscillations seen in the asymmetric samples \cite{Gingrich2016}.

\section{Measurement Procedure}
The samples were initialized into the parallel magnetic state by applying a large (usually -4000 Oe) in-plane magnetic field to align both magnetic layers of both junctions in the negative direction. Although we are only able to measure the relative phase between the two junctions, we interpreted this state as having both junctions in the $\pi$ state based on the arguments outlined in the introduction. Switching the junctions one at a time would then sequentially bring the SQUID from the $\pi-\pi$ state to the $0-\pi$ and then $0-0$ states. After initialization, the chips were briefly raised above the surface of the helium in the cryostat to remove any trapped flux before measurement. IV curves were measured using a self-balancing SQUID-based potentiometer and battery-powered current supply. The flux current $I_\phi$ was provided by a programmable Yokogawa voltage source and a 1 k$\Omega$ resistor. With no applied set field, IV curves were measured as the flux current was swept to identify the location of the central maximum of the SQUID oscillations in the $\pi-\pi$ state. The flux current range was selected to cover several lobes of the critical current oscillations, usually from \SI{-2.5}{\milli\ampere} to \SI{+2.5}{\milli\ampere} for the pitchforks. On the ULI samples, the magnetic field from the flux line suppressed the critical current in the junctions at large flux currents due to the Fraunhofer effect. Near zero field through the junction the critical current remains approximately constant, so we restricted the flux current to a narrow range from \SI{-1}{\milli\ampere} to \SI{+1}{\milli\ampere}. Above this range we saw a systematic decrease in critical current, which we attributed to this Fraunhofer suppression. All transport measurements were taken in zero field to ensure there was no suppression of the critical current from the set field.

\begin{figure}
\includegraphics[width=3.5in]{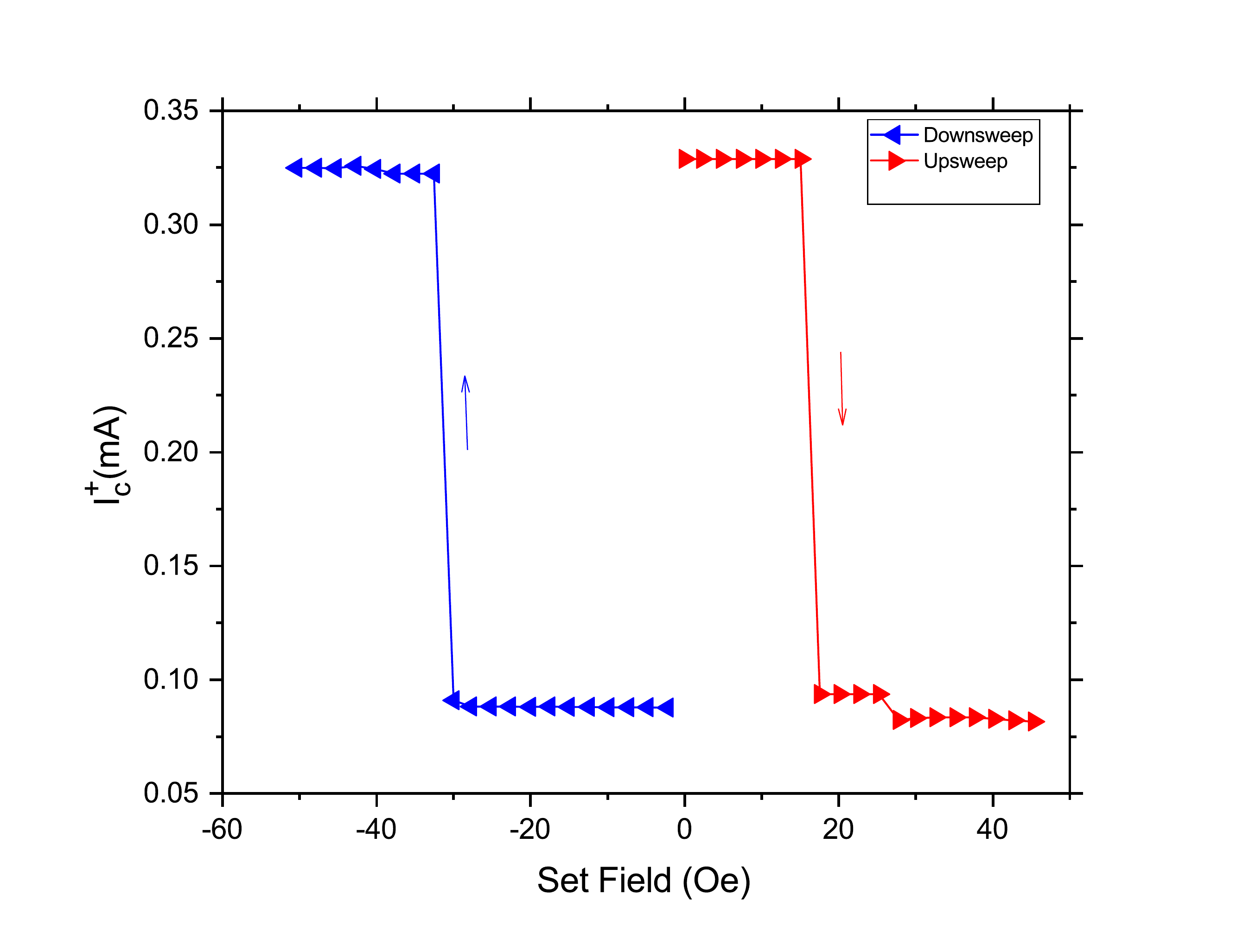}
\caption{Measurement of critical current with flux fixed to identify preliminary values of switching fields for an ultra low inductance SQUID. Sample is initialized in a large field in the negative direction so both junctions are in the parallel state. A flux corresponding to a maximum critical current in the $\pi$ state is applied. The set field is then incremented from zero in the positive direction until a change in the critical current is observed. This change in $I_{c}$ indicates that the free layer in one of the junctions switches at that field. The field is then returned to zero and incremented in negative direction until the device returns to its original state. The field values where the switching occurred are used to select an appropriate field range for the ensuing phase-sensitive measurements. Lines are shown as a guide to the eye.}
\label{ConstantFlux}
\end{figure}

A preliminary measurement to identify the fields necessary to switch the first junction was performed by fixing the flux current at a maximum of the SQUID oscillation and sweeping the set field until a change in critical current is observed, as shown in Figure \ref{ConstantFlux}. The set field is then swept in the negative direction until the junction switches back. We observed a training effect where the switching field varies from run to run in early measurements, so this sweep is repeated until the switching characteristics stabilize from measurement to measurement.
Once the constant-flux characteristics have been identified a full set of phase-sensitive measurements is performed. First, a small set field is applied in-plane. A series of IV curves is then measured in zero set field as the flux current is swept, giving several periods of the critical current oscillation vs flux.  The set field is then incremented, and the process is repeated. Unless otherwise indicated the set field was incremented in steps of 5 Oe. When the first junction switches, an approximately half-period phase shift is observed in the raw data, indicating a switch of one junction from the $\pi$ to zero state. If the set field is further increased (a ``major loop'' measurement) the second junction switches, bringing the phase back to its original value. The set field was never brought above 100 Oe to avoid trapping flux in the devices.

Measurements were performed on SQUIDs on two chips. The first chip had three pitchfork samples, and the second contained four ULI SQUIDs. The large loop inductance of the pitchforks caused a relative shift between $I_{c}^+$ and $I_{c}^-$, as seen in our previous work and described by theory. This offset caused a slanted ``ratchet'' shape in the SQUID oscillations which made it difficult to identify with certainty when a $0-\pi$ transition occurred. Nevertheless, the raw pitchfork data showed a change in the maximum critical current amplitude and a visible phase shift, suggesting that there was some form of switching occurring before extracting the shift in relative phase from a full analysis.

\section{Results}

\begin{figure}[ht]

\includegraphics[width=6in]{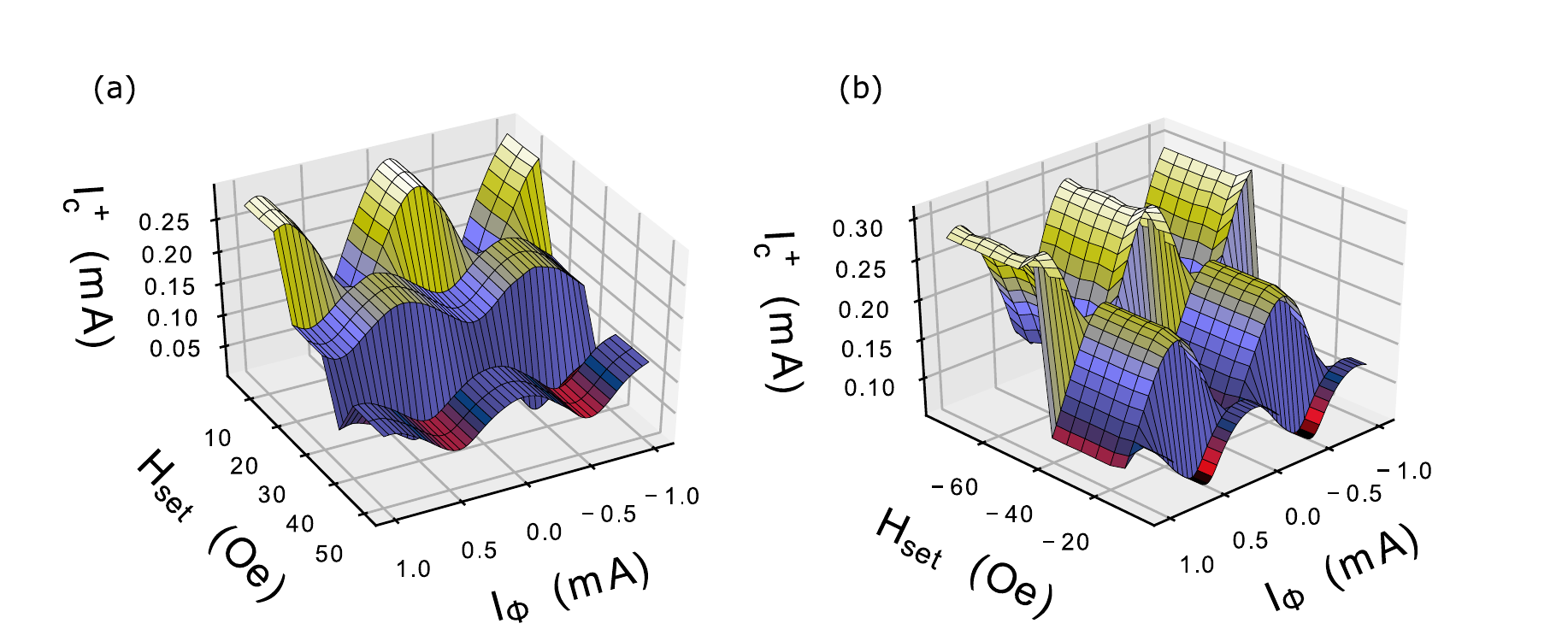}
\caption{3D Plot of critical current I$_c^+$ as a function of set field and flux current for ULI SQUID 4. (a) Major loop upsweep for ULI SQUID. Set field is incremented starting at zero until the free layers in the two junctions switch at 20 and 40 Oe respectively. (b) ULI SQUID swept from 0 to -70 until both junctions both switch back to initialized state shown in (a).}
\label{3D}
\end{figure}

Figure \ref{3D} shows a 3D plot of critical current as a function of flux current and set field for a major loop measurement on an ultra-low inductance sample. Sweeping along the flux current axis shows the usual SQUID oscillation as a function of the flux coupled into the loop. As the set field is swept through the switching fields of the two junctions, the SQUID oscillations show a phase shift and a change in amplitude. Figure \ref{3D}a shows the upsweep data for a ULI SQUID, with the set field being incremented in the positive direction from zero. When the set field reaches 20 Oe in the upsweep, the first junction switches. This is immediately evident in the raw data as the critical current oscillations shift by half a period. There is also a decrease in the maximum critical current as the critical current of these junctions is less in the antiparallel state than parallel. At 40 Oe, the second junction switches. This is again evident by inspection as there is another decrease in critical current and half-period shift. The downsweep data in Figure \ref{3D}b, taken immediately after the upsweep, show the reverse process. At -20 Oe, there is an increase in critical current and half-period shift as the first junction returns to the parallel state. This is repeated at -60 Oe as the second junction switches to the parallel state. The clear half-period shifts make the changes between $0$ and $\pi$ states evident without further analysis.

\begin{figure}
\includegraphics[width=6in]{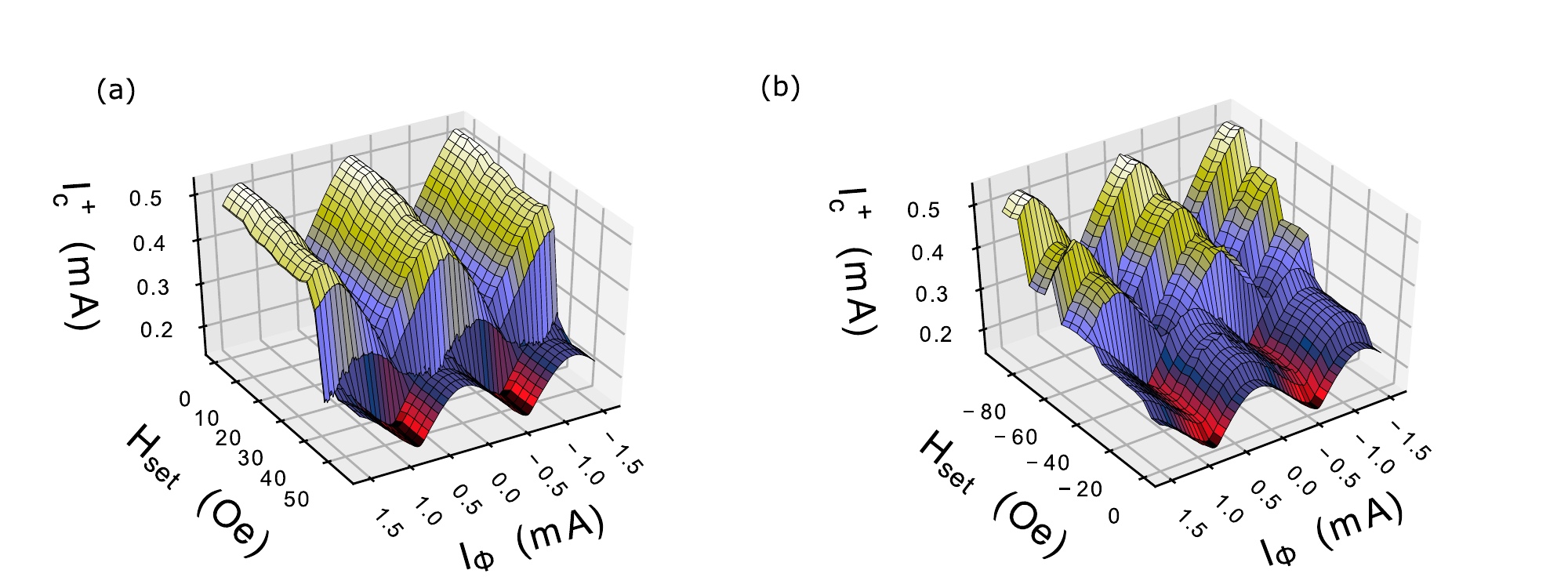}
\caption{Minor loop measurement of critical current vs set field and flux current for pitchfork SQUID 1. (a) Minor loop upsweep for pitchfork sample. One junction switches at 40 Oe, showing a drop in critical current and a half-period phase shift. (b) Downsweep returning pitchfork to initialized state. Amplitude changes in several steps suggest that the free ferromagnetic layer is not switching as a single domain, but the phase shift occurs clearly at -40 Oe.}
\label{3D_Pitchfork}
\end{figure}

Similar data is shown for a pitchfork SQUID in Figure \ref{3D_Pitchfork}. This measurement was a minor loop where the field was kept within a range that would only switch the first junction. The set field is incremented from zero in panel (a) until the first junction switches at 40 Oe. Again, a half-period shift is clearly observed. The set field is then incremented in the negative direction in panel (b) until the junction switches back at -40 Oe. Further changes in the critical current amplitude, but not phase, occur at -60 and -80 Oe, probably due to a domain wall being swept out of the NiFe free layer.

\begin{figure*}[ht]
\includegraphics[width=6in]{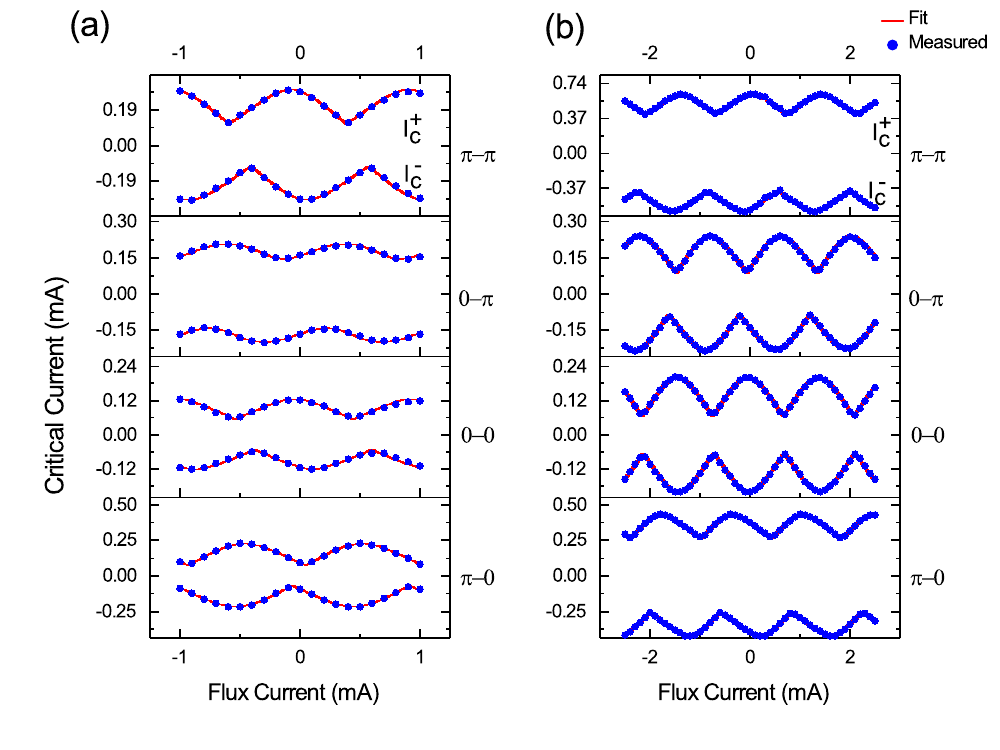}
\caption{(a) Fits (lines) to measured $I_c^+$ and $I_c^-$ data (symbols) in all four states for ULI SQUID 4. A half-period phase change is observed as each junction switches. The first three measurements are cross sections from a major loop in the increasing direction, and the final measurement is from the following downsweep after the first junction switches back to its initial state. The flux current was restricted to \SI{\pm 1}{\milli\ampere} to avoid suppression of the critical current by the Fraunhofer effect. (b) Data and fits to $I_c^+$ and $I_c^-$ in all four states of pitchfork SQUID 2. Differences in critical currents of the two junctions manifest as slight relative shifts between $I_c^+$ and $I_c^-$ in both samples.}\label{FourState}
\end{figure*}

Figure \ref{FourState} shows 2D plots of the critical current vs flux current for the two SQUID designs in the four possible magnetic states, as well as the fits to theory. Panel (a) shows $I_c^+$ and $I_c^-$ for an ultra-low inductance SQUID. The flux current is restricted to a narrow range on the ULI SQUIDs so that the field due to the flux line doesn't significantly suppress the critical current by the Fraunhofer effect.  Panel (b) shows similar data for a pitchfork SQUID. These data show that the $I_c^+ \textrm{ and } I_c^-$ maxima when the junctions are in the same state $(\pi-\pi \textrm{ and } 0-0)$ align with the minima when the junctions are in opposite states $(0-\pi \textrm{ and }\pi-0)$. As the critical current oscillation is periodic in the flux through the SQUID, a $\pi$ phase shift corresponds to a flux shift of $\Phi_0/2$ in the oscillation. The theory fits the data very well and allows the extraction of the flux shift, inductances, and critical currents as described in section \ref{Analysis}. 

\setlength{\tabcolsep}{6pt}
\begin{table*}[hb]
\centering
\label{my-label}
\begin{tabular}{|c|c|c|c|c|c|c|}
\hline
Sample             & State & $L_1$ (pH)    & $L_2$ (pH)  & $I_{c}^1$  (mA)  & $I_{c}^2$ (mA) & $\Delta\Phi_\textrm{shift}/\Phi_0$ \\
\hline
\multirow{2}{*}{1} & $\pi-\pi$   & $2.11 \pm 0.01$ & $2.25 \pm 0.01$ & $0.204 \pm 0.001$ & $0.190 \pm 0.001$ & \multirow{2}{*}{$0.511\pm0.001$} \\ 
                   & $0-\pi$   & $2.17 \pm 0.03$ & $2.32 \pm 0.03$ & $0.204 \pm 0.001$ & $0.114 \pm 0.001$ &  \\ \hline
\multirow{2}{*}{2} & $\pi-\pi$   & $2.22 \pm 0.04$ & $2.26 \pm 0.04$ & $0.122 \pm 0.002$ & $0.173 \pm 0.002$ & \multirow{2}{*}{$0.469\pm0.003$}  \\  
                   & $0-\pi$   & $3.25 \pm 0.13$ & $3.05 \pm 0.12$ & $0.081 \pm 0.002$ & $0.163 \pm 0.002$  & \\ \hline
\multirow{2}{*}{3} & $\pi-\pi$   & $1.93 \pm 0.06$ & $2.09 \pm 0.06$ & $0.174 \pm 0.002$ & $0.191 \pm 0.002$ & \multirow{2}{*}{$0.507\pm0.002$}  \\  
                   & $0-\pi$   & $1.95 \pm 0.05$ & $1.98 \pm 0.05$ & $0.173 \pm 0.001$ & $0.097 \pm 0.00$1 &   \\ \hline
\multirow{4}{*}{4} & $\pi-\pi$   & $2.13 \pm 0.08$ & $2.08 \pm 0.08$ & $0.174 \pm 0.002$ & $0.114 \pm 0.002$ &   \\  
                   & $0-\pi$   & $2.05 \pm 0.52$ & $1.15 \pm 0.42$ & $0.173 \pm 0.001$ & $0.029 \pm 0.001$ &   $0.553\pm0.005$\\ 
                   & $0-0$	   & $1.97 \pm 0.52$ & $1.30 \pm 0.43$ & $0.090 \pm 0.001$ & $0.029 \pm 0.001$ & $-0.508\pm0.008$\\
                   & $\pi-0$   & $1.97 \pm 0.04$ & $2.04 \pm 0.04$ & $0.096 \pm 0.001$ & $0.128 \pm 0.001$ & $0.454\pm0.005$\\ \hline
\end{tabular}
\caption{Overview of best-fit parameters for ULI SQUIDs. Inductances $L_1$ and $L_2$ for the two arms of the SQUID are shown for four samples in the $\pi-\pi$ and  $0-\pi$ states, as well as the critical currents through the two junctions. The SQUID oscillations are periodic in flux, so the shift of the oscillation pattern between states is shown as $\Delta\Phi_\textrm{shift}/\Phi_0$ such that a phase change of $\pi$ would be represented by a $\Phi_0/2$ shift. The critical current of only one junction changes between successive states, consistent with one of the two junctions switching states. There were anomalous changes in the extracted inductances (which should be constant) between states in samples 2 and 4. Data from a major loop was fit for sample 4, showing that each switching event corresponds to a change in one of the two junctions.} 
\end{table*}

Controllable $0-\pi$ switching was observed in all four of the ULI SQUIDS measured in this study. The average phase change for the ULI samples between the $\pi-\pi$ and $0-\pi$ states was $\Delta\Phi_\textrm{shift}/\Phi_0=0.510$. For the pitchforks the average phase change was $\Delta\Phi_\textrm{shift}/\Phi_0=0.529$. We attribute the difference in the pitchfork results to a slight flux being coupled into the SQUID by the NiFe nanomagnets as they switch. This issue is avoided in the ULI design due to their high symmetry.

\begin{table*}[ht]
\centering
\begin{tabular}{|c|c|c|c|c|c|r|}
\hline
Sample             & State & $L_1$ (pH)    & $L_2$ (pH)  & $I_{c}^1$  (mA)  & $I_{c}^2$ (mA) & $\Delta\Phi_\textrm{shift}/\Phi_0$  \\ \hline
\multirow{2}{*}{1} & $\pi-\pi$ & $2.66 \pm 0.03$ & $2.78 \pm 0.03$ & $0.209 \pm 0.00$2 & $0.302 \pm 0.002$ &  \multirow{2}{*}{$0.584 \pm 0.001$}     \\ 
                   & $0-\pi$   & $2.65 \pm 0.03$ & $2.61 \pm 0.03$ & $0.177 \pm 0.001$ & $0.093 \pm 0.001$ &                         \\ \hline
\multirow{4}{*}{2} & $\pi-\pi$ & $2.57 \pm 0.03$ & $2.67 \pm 0.03$ & $0.293 \pm 0.002$ & $0.328 \pm 0.002$ &      \\ 
                   & $0-\pi$   & $2.51 \pm 0.03$ & $2.39 \pm 0.03$ & $0.102 \pm 0.001$ & $0.133 \pm 0.001$ &   $0.439 \pm 0.001$\\
                   & $0-0$	   & $2.47 \pm 0.03$ & $2.32 \pm 0.03$ & $0.102 \pm 0.001$ & $0.099 \pm 0.001$ &   $-0.424 \pm 0.001$\\ 
                   & $\pi-0$   & $2.51 \pm 0.03$ & $2.66 \pm 0.03$ & $0.104 \pm 0.001$ & $0.324 \pm 0.001$ &   $0.468 \pm 0.001$\\ \hline
\multirow{2}{*}{3} & $\pi-\pi$ & $2.91 \pm 0.03$ & $2.90 \pm 0.03$ & $0.248 \pm 0.001$ & $0.249 \pm 0.001$ &  \multirow{2}{*}{$0.564 \pm 0.001$}     \\  
                   & $0-\pi$   & $2.79 \pm 0.07$ & $3.28 \pm 0.07$ & $0.182 \pm 0.001$ & $0.093 \pm 0.001$ &                           \\ \hline
\end{tabular}
\caption{Fit results for pitchfork SQUIDs. The deviation from the expected flux shift of $0.5 \Phi_0$ is attributed to a change in the flux coupled into the SQUID by the NiFe nanomagnets when the free layers switch. Samples 1 and 3 both showed a significant change in critical current for only one junction. Fits are shown for sample 2 at field values along a major loop, where we expected one junction to switch at a time and then one to switch back when the set field was applied in the opposite direction. There is some anomalous behavior observed in the critical current changes for that sample where it appears both junctions are partially switching in the $0-\pi$ state, but flux shifts are still close to $\Phi_0/2$ between states.}
\end{table*}

\section{Analysis}\label{Analysis}
The standard model of a DC SQUID is characterized by four parameters: the inductances of the two arms $L_1$ and $L_2$ and the critical currents of the two junctions $I_{c1}$ and $I_{c2}$. The inductance of the entire SQUID is characterized by the dimensionless parameter $\beta_L \equiv LI_\textrm{c}/\Phi_0$ where $L = L_1+L_2$ is the loop inductance of the SQUID and $I_{c}=I_{c1}+I_{c2}$ is the sum of the critical currents of the two junctions \cite{SQUIDHandbookCh2}. The fits are described by the dimensionless parameters $\alpha_I\equiv \left(I_{c2}-I_{c1}\right)/\left(I_{c2}+I_{c1}\right)$ and $\alpha_L\equiv \left(L_2-L_1\right)/\left(L_2+L_1\right)$ which respectively characterize the asymmetries in the inductances and critical currents between arms. The fit also returns a value of $\Phi_\textrm{shift}$ for each SQUID oscillation. As the period of the oscillation corresponds to one flux quantum, a $\pi$ phase shift in one Josephson junctions induces a horizontal shift in the SQUID oscillation data by $\Phi_0/2$. The phase change was then calculated by taking the difference in $\Phi_\textrm{shift}$ between adjacent states. Because of the symmetric design, a preliminary fit was performed with the inductance asymmetry $\alpha_L$ fixed at zero. The results of this fit were then used as initial guesses for a fit where all four parameters were allowed to vary. For all of the SQUIDs analyzed, the best fit supported half flux quantum shifts between adjacent states, as expected. In the four ULI SQUIDs studied, the fitting program always converged to the expected value of $\Phi_\textrm{shift}$. For the pitchfork SQUIDs, it was possible to force the fit to converge to a value of $\Phi_\textrm{shift}$ differing from the expected value by half a flux quantum by allowing large inductance asymmetries and carefully choosing the initial guess, but the fit consistent with $0-\pi$ switching always had a lower $\chi^2$. As noted in previous work \cite{Gingrich2016}, taking the wrong value for the phase shift also led to unphysical values for the inductances which changed from state to state. We are confident that the fits showing the $\Phi_0/2$ shift incorporate the correct set of parameters for the pitchfork samples, and this issue was not observed in the ultra-low inductance samples.


Inductance simulations were performed in InductEx to compare to the values extracted from the fits for the two SQUID designs. The first round of simulations severely underestimated both the self and mutual inductances of the ULI SQUIDs. As described in section \ref{sec:Fab}, the bottom superconducting electrode consists of a Niobium/Aluminim multilayer instead of pure Niobium. In order for the simulations to acceptably match the experimental data, it was necessary to increase the London penetration depth of the Niobium/Aluminum layer from the conventional Niobium value of \SI{85}{\nano\meter} to \SI{185}{\nano\meter}. For the ULI SQUIDs, the average experimental self inductance (excluding the points that showed unphysically large inductance changes between states)  was $L=\SI{4.10}{\pico\henry}$ and a typical mutual inductance between the flux line and SQUID loop was $M=\SI{2.14}{\pico\henry}$, while the simulated values with $\lambda_L=\SI{185}{\nano\meter}$ were $L=\SI{3.88}{\pico\henry}$ and $M=\SI{2.39}{\pico\henry}$. The experimental results for the pichforks gave $L=\SI{5.33}{\pico\henry}$ and $M=\SI{1.48}{\pico\henry}$. Simulating this design with $\lambda_L=\SI{185}{\nano\meter}$ gave $L=\SI{5.00}{\pico\henry}$ and $M=\SI{1.69}{\pico\henry}$. As a consistency check, we redid the simulations for our asymmetric SQUID study reported in Ref. \cite{Gingrich2016} with $\lambda_L=\SI{185}{\nano\meter}$ for the Niobium/Aluminum layer. The SQUID design in that study differed significantly in that the flux line ran alongside the loop several microns away instead of directly underneath it. Repeating the simulations of that SQUID did not show a significant change in the extracted inductance when $\lambda_L$ was varied.

In our previous work, an increase in maximum critical current was observed as the junctions switched from the parallel to the antiparallel state. In choosing the Ni/NiFe thickness used for this study, the observations shown in Figure 3a of our single-junction results \cite{Niedzielski2018} were used to identify thicknesses where the critical current would be the same in the P and AP states. Those data suggested that for Ni(2.0), a NiFe(1.25) free layer should produce a junction with nearly equal critical currents in the P and AP states. Our data show that we in fact had a decrease in critical current as each junction switched into an AP state, and that in our full stack the thicknesses chosen were not exactly at the crossing point shown in the single-junction data. This characteristic decrease in critical current from the P to AP state is observed for both switches in every sample measured. Given the findings of our single-junction study and the fact that the critical currents we observe are higher in the P state than AP we expect increasing the NiFe thickness slightly would bring the critical currents in the two states closer together.

\section{Conclusions}
In conclusion, we have performed a comprehensive study of SQUIDs containing spin-valve Josephson junctions with a \SI{2.0}{\nano\meter} Ni hard layer and \SI{1.25}{\nano\meter} NiFe soft layer. $0 - \pi$ switching has been observed in these SQUIDs, expanding the phase diagram of known Ni/NiFe thicknesses that support this phase change. This switching has been observed in multiple devices with different designs, showing the fabrication and measurement procedures are robust and repeatable. Introducing the ultra-low inductance SQUID design has greatly simplified analysis, since the phase shift can be directly observed in the raw data.

There is still an extensive amount of work that needs to be done for Ni/NiFe Josephson junctions to be viable for use in a commercial device. While the possibility of phase control is now well-established, there is room for significant improvement in both device design and magnetic characterizations. Over the course of this study the switching field for all of the devices changed significantly from run to run. Several junctions also switched in several steps, rather than the clean switch desired for a binary device. Better understanding the behavior of the ferromagnetic layers and optimizing their switching behavior remains a high priority for future work. 

The amount of data on ferromagnetic layer combinations that can support  $0 - \pi$ switching is still very limited, with prior phase sensitive measurements having been done on Ni(1.2)/NiFe(1.5) \cite{Gingrich2016} and Ni(3.3)/NiFe(1.6) \cite{Dayton2018}. Single-junction measurements \cite{Niedzielski2018} suggest that for Ni(2.0), $0-\pi$ switching should be possible for a range of NiFe thicknesses from about \SI{1.1}{\nano\meter} to \SI{1.5}{\nano\meter}, which is consistent with the findings presented here, and that switching should be possible for Ni(1.6) with NiFe thicknesses between 1.0 and and \SI{1.6}{\nano\meter}. While this study on Ni(2.0)/NiFe(1.25) adds another data point to that phase diagram, there is a significant need for both theoretical advancements in modeling complex spin-valve junctions and broader experimental studies to allow for any sort of optimization of these devices.

These controllable pseudo-spin-valves are part of an increasingly diverse range of Josephson junctions with some form of phase control \cite{Yamashita2018}. This phase control has immediate application in multiple areas of superconducting electronics. A superconducting field-programmable gate array has been designed using magnetic Josephson junctions for using in single-flux-quantum (SFQ) computing \cite{Pedram2018}. $\pi$ junctions have applications in both high-speed low-power classical computing \cite{UstinovKaplunenko2003,Ortlepp2006,Khabipov2010,Kamiya2018} and in quantum computing \cite{Ioffe1999,Blatter2001,Yamashita2005,Feofanov2010}. The ability to control the phase state of a junction in-situ will surely lead to a wealth of new circuit designs. 

\begin{acknowledgments}
We are grateful to V. Aguilar for his assistance with measurement software and instrumentation. We are grateful to B.M. Niedzielski and J.A. Glick for their support with sample fabrication and measurement. We acknowledge valuable discussion with E.C. Gingrich, D. Miller, N.D. Rizzo, O. Naaman, D.S. Holmes, and A.Y. Herr. We also thank B. Bi for technical assistance and the use of the W.M. Keck Microfabrication Facility. This research is based upon work supported by the ODNI, IARPA, via ARO contract number W911NF-14-C-0115. The views and conclusions contained herein are those of the authors and should not be interpreted as necessarily representing the official policies or endorsements, either expressed or implied, of the ODNI, IARPA, or the U.S. Government.
\end{acknowledgments}

\section*{References}
\bibliographystyle{unsrt}
\bibliography{Madden_Spin_Valve_Bib}
\end{document}